# Achievable Rates of MIMO Systems with Linear Precoding and Iterative LMMSE Detection

Xiaojun Yuan, *Member, IEEE*, Li Ping, *Fellow, IEEE*, Chongbin Xu, and Aleksandar Kavcic, *Senior Member, IEEE*

*Abstract*—We establish area theorems for iterative detection over coded linear systems (including multiple-input multiple-output (MIMO) channels, inter-symbol-interference (ISI) channels, and orthogonal frequency-division multiplexing (OFDM) systems). We propose a linear precoding technique that asymptotically ensures the Gaussianness of the messages passed in iterative detection, as the transmission block length tends to infinity. We show that the proposed linear precoding scheme with iterative linear minimum mean-square error (LMMSE) detection is potentially information lossless, under various assumptions on the availability of channel state information at the transmitter (CSIT). Numerical results are provided to verify our analysis.

## I. Introduction

### A. Area Properties

The optimal detection of a coded signal in a complicated channel environment may incur excessive complexity. Iterative detection provides a low-cost solution by decomposing the overall receiver into two or more local processors (cf., [1]-[11] and the references therein). The analysis of an iterative detection process is an intriguing problem. The density-evolution technique [2] shows that carefully designed low-density parity-check (LDPC) codes can achieve near-capacity performance in additive white Gaussian noise (AWGN) channels with iterative message-passing decoding algorithms. It was further shown in [12] that the achievable rate of an iterative scheme for an erasure channel can be measured by the area under the so-called extrinsic information transfer (EXIT) curves [13] and the channel capacity is approachable when the two local processors have matched EXIT curves. This area property is extended in [14] to scalar AWGN channels (or simply, AWGN channels) using the measure of minimum mean-square error (MMSE), which establishes a sufficient condition to approach the capacity of a binary-input AWGN channel with iterative detection.

It is commonly accepted [13] that, with random interleaving, the extrinsic information (i.e., the messages) from an *a posteriori* probability (APP) decoder for a binary forward-error-control code can be modeled as a sequence of observations from an effective AWGN channel. Thus, the authors in [10] made two basic assumptions: (i) the messages passed between the local processors are modeled as the observations from an effective AWGN channel (referred to as the AWGN assumption) and (ii) the local processors are optimal in the sense of APP detection/decoding. Assumption (i) ensures that each local processor can be characterized by a single-variable transfer function involving signal-to-noise ratio (SNR) and minimum mean-square error (MMSE). The mutual information and MMSE relationship established in [12] can then be applied to derive the area property when the transfer curves of the local processors are matched.

### B. MIMO Channels

The work in [14] is for scalar channels. Its extension to multiple-input multiple-output (MIMO) channels is not straightforward. For example, the signals sent from different transmit antennas in a MIMO channel may experience different channel conditions. Multivariate functions can be used to characterize the receiver's behavior, but then assumption (i) mentioned above may not hold.

Also, the optimal APP detection for a local processor may be costly in a MIMO environment. Linear MMSE (LMMSE) detection [18] is an attractive low-complexity alternative, but it is suboptimal for non-Gaussian signaling, implying that assumption (ii) may not hold.

### C. Contributions of This Paper

In this paper, we consider a joint linear precoding (LP) and iterative LMMSE detection scheme. We show that the proposed LP technique ensures that the AWGN assumption asymptotically holds for the output of the LMMSE detector, provided that the transmission block length is sufficiently large. This allows us to use a single pair of input output parameters to characterize the behavior of the LMMSE detector. Also, we adopt superposition coded modulation (SCM) [19][20] which leads to approximately Gaussian signaling, for which LMMSE detection is near-optimal. Based on that, we establish area theorems for the proposed LP and iterative LMMSE detection (LP-LMMSE) scheme in MIMO channels, under various assumptions on the availability of channel state information at the transmitter (CSIT). We show that the proposed LP-LMMSE scheme is potentially information lossless based on the curve-matching principle, even though the suboptimal iterative LMMSE detection technique is employed.

### D. Comparisons with Existing LP Techniques

It is interesting to compare the proposed linear precoding technique with other alternatives (cf., [21]-[28] for full CSIT and [29]-[34] for partial CSIT). From an information-theoretic viewpoint, the ultimate criterion for precoder design is to achieve the channel capacity (cf., [23][33][34] and the references therein). It is well known that channel coding is required to achieve the capacity. However, most existing works on precoder design are focused on un-coded MIMO systems equipped with a signal detector at the receiver.

Along this line, a variety of design criteria have been studied, such as pair-wise error probability (PEP) minimization [22], minimum signal-to-interference-plus-noise ratio (SINR) maximization (min-max) [23], average SINR maximization [23], SINR equalization [24][25], and MMSE [26][27], etc. The works for un-coded systems can be applied to coded systems by concatenating the detector with a decoder. Unfortunately, the residual interference at the output of the detector may result in a considerable performance loss [36][55][56].

Iterative detection and decoding can efficiently suppress the residual interference left by the detector. Then, a major challenge is to jointly design channel coding and linear precoding at the transmitter, taking into account the effect of iterative detection at the receiver. This paper provides a simple solution to this problem. Our analysis shows that, with curve-matching codes, the proposed LP-LMMSE scheme is capacity-achieving under various assumptions on CSIT.

## II. SYSTEM MODEL AND ITERATIVE LMMSE DETECTION

### A. Generic Linear System

A generic complex-valued linear system is modeled as

$$y = Ax + \eta \quad (1)$$

where $y$ is a received signal vector, $A$ is a transfer matrix, $x$ is a transmit signal vector, and $\eta \sim \mathcal{CN}(0, \sigma^2 I)$ an additive noise vector. Notation $0$ (or $I$) represents an all-zero vector (or an identity matrix) with a proper size. Here we only specify that the length of the signal vector $x$ is denoted by an integer $J$. The sizes of the other matrices and vectors above will be revealed later.

Throughout this paper, we assume full channel state information (CSI) at the receiver, i.e., the receiver perfectly knows the transfer matrix $A$. We will discuss the situations for both perfect and partial CSIT.

### B. Messages

The transmitter structure for the system in (1) is shown in the upper part of Fig.1. The encoder generates a frame of $KJ$ coded symbols (denoted by $x'$), where $K$ is the number of system uses. These coded symbols are randomly interleaved by the interleaver $\Pi$ and then partitioned into $K$ segments with equal length. Each segment (consisting of $J$ symbols) serves as an input $x$ to the system (1).

The iterative receiver, as illustrated in the lower half of Fig.1, consists of two local processors, namely, the detector and the decoder, inter-connected by the interleaver $\Pi$ and the de-interleaver $\Pi^{-1}$. Particularly, the detector considered in this paper follows the LMMSE principle, hence the name LMMSE estimator.

The LMMSE estimator estimates $x$ based on the channel observation $y$ and the messages from the decoder (denoted by $\alpha$). The outputs of the estimator are extrinsic messages (denoted by $\beta$). The message set $\alpha$ is defined as follows. Denote by $x_i$ the $i$th entry of $x$. Each $x_i$ is constrained on a discrete signaling constellation $\mathcal{S} = \{s_1, s_2, ..., s_{|\mathcal{S}|}\}$, where $|\mathcal{S}|$ represents the size of $\mathcal{S}$. For each use of the system (1), $\alpha$ is a set of $J$ messages as

$$\alpha = \{\alpha_1, \alpha_2, ..., \alpha_J\},$$

and each message $\alpha_i$ is a set of $|\mathcal{S}|$ likelihood values for $x_i$, i.e.

$$\alpha_i = \{\alpha_i(1), \alpha_i(2), ..., \alpha_i(|\mathcal{S}|)\},$$

where $\alpha_i(k)$ represents the likelihood of $x_i = s_k \in \mathcal{S}$ prior to the processing of the detector, and

$$\sum_{k=1}^{|\mathcal{S}|} \alpha_i(k) = 1.$$

Similar notation applies to the message set $\beta$.

The decoder decodes $x'$ based on the input message set $\beta'$, and outputs the extrinsic message set $\alpha'$. Decoding is performed on the overall frame, and so both $\alpha'$ and $\beta'$ contain $KJ$ messages. Let $x'_i$ be the $i$th entry of $x'$, and $\beta'_i(k)$ be the likelihood of $x'_i = s_k \in \mathcal{S}$ prior to decoding. Then, we can express

$$\beta' = \{\beta'_i(k) \mid i = 1, ..., KJ \text{ and } k = 1, ..., |\mathcal{S}|\}.$$

Messages in $\alpha'$ allow similar expressions. After decoding, $\alpha'$ are interleaved and partitioned to form the input of the LMMSE estimator, which completes one round of iteration. The LMMSE estimator and the decoder are executed iteratively until convergence.

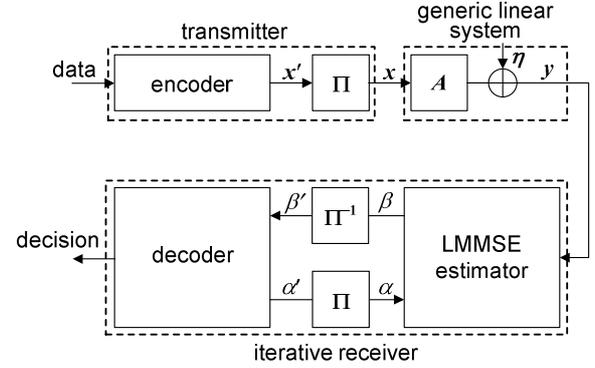

Fig. 1. The transceiver structure of the proposed scheme over the generic linear system in (1). $\Pi$ and $\Pi^{-1}$ are two complementary interleavers.

### C. Basic Assumptions

Here we discuss some basic assumptions used throughout this paper. For notational convenience, we introduce an auxiliary random variable, denoted by $a_i$, to represent the information carried by each $\alpha_i$. The conditional probability of $a_i$ given $x_i$ is defined as

$$p(a_i \mid x_i = s_k) = \alpha_i(k).$$

Similarly, an auxiliary random variable $b'_i$ is defined for each $\beta'_i$. The conditional probability of $b'_i$ given $x_i$ is given by

$$p(b'_i \mid x_i = s_k) = \beta'_i(k).$$

We make the following assumptions on the inputs of the local processors.

*Assumption 1:* For the detector, each $x_i$ is independently drawn from $\mathcal{S}$ with equal probability, i.e., $p(x_i = s_k) = 1/|\mathcal{S}|$, for any $i$ and $k$;[1] the messages $\{a_i\}$ are conditionally independent and identically distributed (i.i.d.) given $x$, i.e.

$$p(\boldsymbol{a} \mid \boldsymbol{x}) = \prod_{i=1}^{J} p(a_i \mid x_i),$$

---

[1] Strictly speaking, $\{x_i\}$ are correlated as $x'$ is coded.

and for any dummy variable $t$ and $k = 1, 2, \ldots, |\mathcal{S}|$,
$$p(a_i = t \mid x_i = s_k) = p(a_j = t \mid x_j = s_k).$$

*Assumption 2:* For the decoder, $\{b_i'\}$ are conditionally i.i.d. given $\mathbf{x}'$, i.e.
$$p(\mathbf{b}' \mid \mathbf{x}) = \prod_{i=1}^{JK} p(b_i' \mid x_i),$$
and for any dummy variable $t$ and $k = 1, 2, \ldots, |\mathcal{S}|$,
$$p(b_i' = t \mid x_i = s_k) = p(b_j' = t \mid x_j = s_k).$$

With Assumption 1, the joint probability space of $\mathbf{x}$, $\mathbf{a}$, and $\mathbf{y}$ seen by the detector can be expressed as
$$p(\mathbf{x}, \mathbf{a}, \mathbf{y}) = \left( \prod_{i=1}^{J} p(x_i) p(a_i \mid x_i) \right) p(\mathbf{y} \mid \mathbf{x}), \quad (2a)$$
where $p(\mathbf{y}|\mathbf{x})$ is determined by (1). Similarly, with Assumption 2, the joint probability space of $\mathbf{x}'$ and $\mathbf{b}'$ seen by the decoder is
$$p(\mathbf{x}', \mathbf{b}') = p(\mathbf{x}') \prod_{i=1}^{JK} p(b_i' \mid x_i'). \quad (2b)$$

Assumptions 1 and 2 decouple the probability space seen by the detector and decoder, which simplifies the analysis of the iterative process. Similar assumptions have been widely used in turbo/iterative detection algorithms [2][13][17][37]. These assumptions are asymptotically ensured by random interleaving as $K$ tends to infinity. In what follows, we always assume that $K$ is sufficiently large so that Assumptions 1 and 2 holds.

*D. LMMSE Estimation*

The detector delivers the *extrinsic* messages defined as
$$\beta_i(k) = p(x_i = s_k \mid \mathbf{a}_{-i}, \mathbf{y})$$
for $i = 1, 2, \ldots, J$ and $k = 1, \ldots, |\mathcal{S}|$, where the joint probability space of $\mathbf{x}$, $\mathbf{a}$, and $\mathbf{y}$ is given in (2a), and $\mathbf{a}_{-i}$ represents the vector obtained by deleting the $i$th entry of $\mathbf{a}$. Here, "extrinsic" means that the contribution of the *a priori* message $a_i$ is excluded in calculating each $\beta_i$.

Direct evaluation of $\beta_i(k)$ above can be excessively complicated for system (1). The LMMSE estimation is a low cost alternative. It is suboptimal in general, but it is optimal if $\mathbf{x}$ are generated using Gaussian signaling.

Denote the mean and covariance of $\mathbf{x}$ (seen by the detector) by $\bar{\mathbf{x}} = [\bar{x}_1, \ldots, \bar{x}_J]^T$ and $v\mathbf{I}$, respectively, with
$$\bar{x}_i = \mathrm{E}[x_i \mid a_i] = \sum_{k=1}^{|\mathcal{S}|} \alpha_i(k) s_k \quad (3a)$$
and
$$v = \mathrm{E}\left[ |x_i - \bar{x}_i|^2 \right], \text{ for any index } i, \quad (3b)$$
where the expectation is taken over the joint distribution of $a_i$ and $x_i$. From Assumption 1, $v$ is invariant with respect to the index $i$. In practice, $v$ is approximated by the sample variance as
$$v \approx J^{-1} \sum_{i=1}^{J} \sum_{k=1}^{|\mathcal{S}|} \alpha_i(k) |s_k - \bar{x}_i|^2,$$
which converges to the true variance when $J \to \infty$.

The LMMSE estimator of $\mathbf{x}$ given $\mathbf{y}$ is [38]
$$\hat{\mathbf{x}} = \bar{\mathbf{x}} + v \mathbf{A}^H \mathbf{R}^{-1} (\mathbf{y} - \mathbf{A}\bar{\mathbf{x}}) \quad (4a)$$
where $\mathbf{R}$ is the covariance matrix of $\mathbf{y}$ given as
$$\mathbf{R} = v \mathbf{A} \mathbf{A}^H + \sigma^2 \mathbf{I}. \quad (4b)$$
Recall that the extrinsic message $\beta_i$ should be independent of $a_i$. To meet this requirement, we calculate the extrinsic mean and variance for $x_i$ (denoted by $b_i$ and $u_i$, respectively) by excluding the contribution of $a_i$ according to the Gaussian message combining rule (cf., (54) and (55) in [39]) as
$$u_i^{-1} = M(i,i)^{-1} - v^{-1} \quad (5a)$$
and
$$\frac{b_i}{u_i} = \frac{\hat{x}_i}{M(i,i)} - \frac{\bar{x}_i}{v}, \quad (5b)$$
where $M(i, i)$ represents the $(i, i)$th entry of the MMSE matrix
$$\mathbf{M} = v\mathbf{I} - v^2 \mathbf{A}^H \mathbf{R}^{-1} \mathbf{A}. \quad (5c)$$
Finally, the output messages of the detector can be calculated as
$$\beta_i(k) = c \cdot p(b_i \mid x_i = s_k), \text{ for any } i \text{ and } k \quad (5d)$$
where $c$ is a scaling factor to ensure
$$\sum_{k=1}^{|\mathcal{S}|} \beta_i(k) = 1.$$
In the above, $\beta_i(k)$ can be readily calculated by assuming that $b_i$ is an observation of $x_i$ over an effective AWGN channel with noise power $u_i$. The justification of this assumption can be found in Lemma 1 in Section III.C.

*E. APP Decoding*

The decoder decodes $\mathbf{x}'$ based on the messages $\boldsymbol{\beta}'$ following the *a posteriori* probability (APP) decoding principle. The extrinsic output of the decoder for each $x_i'$ is defined as
$$\alpha_i'(k) = p(x_i' = s_k \mid \mathbf{b}_{-i}') \quad (6)$$
for $i = 1, 2, \ldots, KJ$ and $k = 1, \ldots, |\mathcal{S}|$, where the joint probability space of $\mathbf{x}'$ and $\mathbf{b}'$ is given in (2b).

*Assumption 3:* The local decoder performs APP decoding.

In practice, APP decoding is usually computationally involving. Low-complexity message-passing algorithms can be used to achieve near-optimal performance, provided that the code structure allows a sparse graphic description [37]. Message-passing decoding is well-studied in the literature, and thus the details are omitted here. Compared with APP decoding, the performance loss of message-passing decoding is usually marginal. This loss is not of concern in this paper. Therefore, we introduce Assumption 3 to simplify our analysis.

## III. LINEAR PRECODING WITH PERFECT CSIT

In this section, we assume perfect CSIT. The case of imperfect CSIT will be discussed in Section V. We propose a linear precoding technique to equalize the output SINRs of the detector. We then establish an SINR-variance transfer chart technique to analyze the performance of the proposed LP-LMMSE scheme.

*A. MIMO Channels*

A Gaussian MIMO channel with $N$ transmit and $M$ receive antennas can be modeled as
$$\tilde{\mathbf{y}}_i = \mathbf{H} \tilde{\mathbf{x}}_i + \tilde{\boldsymbol{\eta}}_i \quad (7)$$
where $i$ represents the $i$th channel use, $\tilde{\mathbf{y}}_i$ is an $M$-by-1 received signal vector, $\mathbf{H}$ is the $M$-by-$N$ channel transfer matrix known at both the transmitter and receiver, $\tilde{\mathbf{x}}_i$ is an $N$-by-1 transmit signal vector, and $\tilde{\boldsymbol{\eta}}_i \sim \mathcal{CN}(\mathbf{0}, \sigma^2 \mathbf{I})$ is an $M$-by-1 additive noise vector.

A transmission block (i.e., one use of system (1)) involves $J/N$ uses of the channel (7), where $J$ is the block length defined in Section II.A. We assume that $J$ is properly chosen so that $J/N$ is an integer. Denote

$$\tilde{y} = [\tilde{y}_1^T, ..., \tilde{y}_{J/N}^T]^T$$
$$\tilde{x} = [\tilde{x}_1^T, ..., \tilde{x}_{J/N}^T]^T$$
$$\tilde{\eta} = [\tilde{\eta}_1^T, ..., \tilde{\eta}_{J/N}^T]^T .$$

Combining $J/N$ channel uses, we write an extended system as

$$\tilde{y} = \tilde{H}\tilde{x} + \tilde{\eta} \quad (8a)$$

where the extended channel is given by

$$\tilde{H} = \begin{bmatrix} H & & & \\ & H & & \\ & & \ddots & \\ & & & H \end{bmatrix}. \quad (8b)$$

The channel input is power constrained as

$$E[\tilde{x}^H \tilde{x}]/J \leq P. \quad (8c)$$

The singular value decomposition (SVD) of $\tilde{H}$ is given by

$$\tilde{H} = U\Lambda V^H \quad (9)$$

where $\Lambda$ is an $(JM/N)$-by-$J$ diagonal matrix with non-negative diagonal elements, and $U$ and $V$ are unitary matrices. We assume that $U$ and $V$ are chosen such that the diagonal entries of $\Lambda$ are asymptotically uncorrelated as $J$ tends to infinity.[2] This property is useful in establishing Lemma 1 in Subsection C.

*B. Linear Precoding*

We focus on the following linear precoding operation:

$$\tilde{x} = VW^{1/2}Fx \quad (10a)$$

where $\tilde{x}$ and $x$ are given in (8a) and (1), respectively, $V$ is defined in (9), $W$ is a diagonal matrix with non-negative diagonal elements for power allocation, and $F$ is the normalized DFT matrix with the $(i, k)$th entry given by

$$F(i, k) = J^{-1/2}\exp(-j2\pi(i-1)(k-1)/J) \quad (10b)$$

with $j = \sqrt{-1}$.

From Assumption 1, the entries of $x$ are uncorrelated. Without loss of generality, we further assume that the entries of $x$ have normalized power. Denote the diagonal entries of $W$ by $\{W(1,1), ..., W(J,J)\}$. Then the power constraint in (8c) becomes

$$J^{-1}\sum_{i=1}^{J} W(i,i) \leq P. \quad (11)$$

The optimization of $W$ will be detailed in Section IV.

The use of $F$ in (10a) is to ensure that the SINRs are equal for all symbols after LMMSE detection. (See the discussions in Subsection C.) Incidentally, the choice of the DFT matrix for $F$ also allows the fast Fourier transform (FFT) algorithm in the implementation of LMMSE detection [10]. We also note that similar DFT-based precoders have been used in MIMO systems for other purposes, e.g., for harnessing diversity in [60].

At the receiver side, the received vector $r$ is post-processed by the matrix $U^H$:

$$y = U^H \tilde{y}. \quad (12)$$

Combining (8)-(12) and letting

$$D = \Lambda W^{1/2}, \quad (13a)$$

we obtain an equivalent channel as

$$y = DFx + U^H\tilde{\eta} = Ax + \eta \quad (13b)$$

where $A = DF$ and $\eta = U^H\tilde{\eta} \sim \mathcal{CN}(0, \sigma^2 I)$ since $U$ is unitary. Clearly, (13b) has the same form as (1). The iterative detection and decoding procedure outlined in Section II can be directly applied to (13b).

*C. Characterization of the Estimator*

It was shown in [23] that, with the precoder in (10a), the SINR becomes uniform with (non-iterative) LMMSE estimation. We will see that a similar property holds in our proposed iterative system. Note that the precoder in (10a) is applied to the extended system in (8b). As we will see, for the extended system, the residual interference and noise term in the output of the estimator is asymptotically Gaussian following the central limit theorem.

To start with, define

$$\phi(v) \triangleq \frac{1}{M(i,i)} - \frac{1}{v}.$$

This definition is based on the fact that, $M(i, i)$ is not a function of $i$. To see this, substitute (4b) and $A = DF$ into (5c):

$$M = vI - v^2 F^H D^H (vDD^H + \sigma^2 I)^{-1} DF.$$

Then

$$M(i,i) = v - v^2 f_i^H D^H (vDD^H + \sigma^2 I)^{-1} Df_i \quad (14a)$$

$$= v - J^{-1}\sum_{k=1}^{J} \frac{v^2 D(k,k)^2}{vD(k,k)^2 + \sigma^2} \quad (14b)$$

$$= \frac{1}{J}\sum_{k=1}^{J}\left(v^{-1} + \frac{D(k,k)^2}{\sigma^2}\right)^{-1} \quad (14b)$$

where $f_i$ is the $i$th column of $F$, and (14b) utilizes (10b). Clearly, $M(i, i)$ is the same for all $i$. Furthermore, recall that $\{D(k, k) = \Lambda(k, k)W(k, k)^{1/2}, k = 1, 2, ..., J\}$ and that $\{\Lambda(k, k)\}$ are the singular values of $\tilde{H}$ in (8b) and so $\{\Lambda(k, k)\}$ contains $J/N$ copies of the singular values of $H$ (denoted by $\lambda_1, ..., \lambda_N$). We assume that a same amount of power is allocated to any two eigen-modes with the same channel gains, i.e., if $\Lambda(k, k) = \Lambda(j, j)$, then $W(k, k) = W(j, j)$.[3] Then, $\{D(k, k)\}$ have at most $N$ different values, denoted by $d_1, ..., d_N$. Using (14b) and the definition of $\phi(v)$, we obtain

$$\rho = \phi(v) = \left(\frac{1}{N}\sum_{n=1}^{N}\left(\frac{1}{v} + \frac{d_n^2}{\sigma^2}\right)^{-1}\right)^{-1} - v^{-1}. \quad (15)$$

Next, we show that the outputs of the LMMSE estimator can be model as the observations from an AWGN channel provided that $J$ is sufficiently large, and the related SINR is given by $\rho = \phi(v)$. Define

$$n_i \triangleq \frac{v}{M(i,i)\rho} f_i^H D(vD^H D + \sigma^2 I)^{-1}(A(x_{\setminus i} - \bar{x}_{\setminus i}) + \eta) \quad (16)$$

where $x_{\setminus i}$ (or $\bar{x}_{\setminus i}$) represents the vector obtained by setting the $i$th entry of $x$ (or $\bar{x}$) to zero. It is clear that $n_i$ is independent of $x_i$. Moreover, we have the following result with the proof given in Appendix A.

---

[2] Although $\tilde{H}$ is deterministic, the ordering of its singular values can be arbitrarily chosen. Here we choose a series of orderings (one for each block length $J$) such that the diagonal of $\Lambda$ is asymptotically uncorrelated as $J$ tends to infinity.

[3] This treatment doesn't incur any information loss, as seen from Theorem 1.

*Lemma 1:* For any index $i$, $b_i$ in (5b) can be expressed as
$$b_i = x_i + n_i, \quad (17)$$
where $n_i$ is independent of $x_i$ and its distribution converges to $\mathcal{CN}(0, 1/\phi(v))$ as $J \to \infty$.

In the above, $n_i$ represents the residual interference plus noise at the output of the LMMSE estimator in iteration. Hence, $\rho = \phi(v)$ in (15) represents the related SINR. It is well-known that the residue error of the LMMSE estimation is approximately Gaussian [38][52]. In Appendix A, we show that this approximation becomes exact for the precoder in (10a) over the extended channel in (8b).

We henceforth assume that $J$ is sufficiently large, so that $\{b_i\}$ can be modeled as the observations of $\{x_i\}$ from an effective AWGN channel with SNR = $\phi(v)$.

*D. Characterization of the Decoder*

We next consider the characterization of the decoder's behavior. The decoder performs APP decoding upon receiving the messages $\boldsymbol{b}' = \{b_i'\}$ modeled as independent observations of $\boldsymbol{x}' = \{x_i'\}$ over an AWGN channel with SNR = $\rho$ (cf., Lemma 1). Define the extrinsic variance of each $x_i'$ as
$$\mathrm{MMSE}(x_i' \mid \boldsymbol{b}_{-i}') = \mathrm{E}\left[|x_i' - \mathrm{E}[x_i' \mid \boldsymbol{b}_{-i}']|^2\right], \quad (18a)$$
where the expectation is taken over the joint probability space of $\boldsymbol{x}'$ and $\boldsymbol{b}'$. We assume that the channel code is properly constructed so that the above MSE is invariant with respect to the index $i$. Then, we can express this MSE (denoted by $v$) as a function of $\rho$:
$$v = \psi(\rho) \quad (18b)$$
referred to as the SNR-variance transfer function of the decoder.

We next establish a relation between the code rate and $\psi(\rho)$. To this end, define the MMSE of each $x_i'$ after decoding as
$$mmse(\rho) = \mathrm{MMSE}(x_i' \mid \boldsymbol{b}')$$
$$= \mathrm{E}\left[|x_i' - \mathrm{E}[x_i' \mid \boldsymbol{b}']|^2\right], \quad (19)$$
where the expectation is taken over the joint probability space of $\boldsymbol{x}'$ and $\boldsymbol{b}'$. Compared with (18), the MMSE in (19) is obtained by including the contribution of $b_i'$ in estimating $x_i'$.

To proceed, we establish a connection between $\psi(\rho)$ and $mmse(\rho)$. We first assume that (i) $x_i'$ is Gaussian, and that (ii) the extrinsic messages are Gaussian. Then, $mmse(\rho)$ and $\psi(\rho)$ are related as
$$mmse(\rho) = \mathrm{MMSE}(x_i' \mid \boldsymbol{b}')$$
$$= \mathrm{MMSE}(x_i' \mid b_i', \boldsymbol{b}_{-i}')$$
$$\overset{(a)}{=} \mathrm{MMSE}(x_i' \mid b_i', x_i' \sim p(x_i' \mid \boldsymbol{b}_{-i}'))$$
$$\overset{(b)}{=} \mathrm{MMSE}(x_i' \mid b_i', x_i' \sim \mathcal{CN}(\mathrm{E}[x_i' \mid \boldsymbol{b}_{-i}'], \psi(\rho)))$$
$$\overset{(c)}{=} (\rho + \psi(\rho)^{-1})^{-1}. \quad (20)$$
where step $(a)$ follows from the fact that $b_i' \to x_i' \to \boldsymbol{x}_{-i}' \to \boldsymbol{b}_{-i}'$ forms a Marcov chain (from Assumption 2 in Section II.C), step $(b)$ from Assumptions (i) and (ii) that
$$p(x_i' \mid \boldsymbol{b}_{-i}') = \mathcal{CN}(\mathrm{E}[x_i' \mid \boldsymbol{b}_{-i}'], \psi(\rho))),$$
and step $(c)$ utilizes the fact that $b_i'$ can be modeled as an AWGN observation in Lemma 1. From [14] (cf., Lemma 1 in [14]), we express the code rate (per symbol of $\boldsymbol{x}'$) as
$$R = \int_0^\infty mmse(\rho)\,d\rho = \int_0^\infty (\rho + \psi(\rho)^{-1})^{-1}\,d\rho. \quad (21)$$

Eqn. (21) is the relation we intend to establish. However, the above derivation is based on two unjustified assumptions (namely, assumptions (i) and (ii)). These assumptions are difficult to meet exactly, as almost all practical systems employ discrete signaling. Intuitively, superposition coded modulation (SCM) [19][20] can be used to approximate Gaussian signaling. We next show that the rate in (21) is indeed approachable by properly constructing an SCM-based channel code. Our result is summarized in the lemma below.

*Lemma 2:* Assume that an arbitrary function $\psi(\rho)$ is first-order differentiable and monotonically decreasing in $\rho$, and
$$\lim_{\rho \to \infty} \rho \psi(\rho) = 0. \quad (22)$$
Let $\Gamma_n$ be an $n$-layer SCM code with SINR-variance transfer function $\psi_n(\rho)$ and rate $R_n$. Then, there exist $\{\Gamma_n\}$ such that: (i) $\psi_n(\rho) \le \psi(\rho)$, for any $\rho \ge 0$ and any integer $n$; (ii) as $n \to \infty$,
$$R_n \to \int_0^\infty (\rho + \psi(\rho)^{-1})^{-1}\,d\rho.$$

The proof of Lemma 2 is given in Appendix B. This lemma reveals that the rate in (21) is indeed approachable.

*E. SINR-Variance Transfer Chart*

From the previous discussions, the LMMSE estimator can be characterized by $\rho = \phi(v)$; and similarly, the decoder can be characterized by $v = \psi(\rho)$. The iterative process of the estimator and decoder can be tracked by a recursion of $\rho$ and $v$. Let $i$ be the iteration number. We have
$$\rho^{(i)} = \phi(v^{(i-1)}) \text{ and } v^{(i)} = \psi(\rho^{(i)}), \, i = 1, 2, \ldots$$
The recursion continues and converges to a point $v^*$ satisfying
$$\phi(v^*) = \psi^{-1}(v^*) \text{ and } \phi(v) > \psi^{-1}(v), \text{ for } v \in (v^*, 1]$$
where $\psi^{-1}(\cdot)$ is the inverse of $\psi(\cdot)$ which exists since $\psi(\cdot)$ is continuous and monotonic (cf., [43]). Note that $v \le 1$ since the signal power is normalized; $v^* = 0$ implies that $\boldsymbol{x}$ can be perfectly recovered. The above recursive process is illustrated by the SINR-variance transfer chart in Fig.2.

We say that the detector and the decoder are matched if
$$\phi(v) = \psi^{-1}(v), \text{ for } v \in (0, 1]. \quad (23)$$
Note that: $\phi(1) > 0$ since the detector's output always contains the information from the channel even if there is no information from the decoder; and $\phi(0) < \infty$ since the detector's extrinsic output cannot resolve the uncertainty introduced by the channel noise even if the messages from the decoder are perfectly reliable. Then, we equivalently express the curve-matching condition (23) as
$$\psi(\rho) = \phi^{-1}(\phi(1)) = 1, \text{ for } 0 \le \rho < \phi(1); \quad (24a)$$
$$\psi(\rho) = \phi^{-1}(\rho), \quad \text{for } \phi(1) \le \rho < \phi(0); \quad (24b)$$
$$\psi(\rho) = 0, \quad \text{for } \phi(0) \le \rho < \infty. \quad (24c)$$

The above curve-matching principle plays an important role in establishing the area theorems, as seen in the next section.

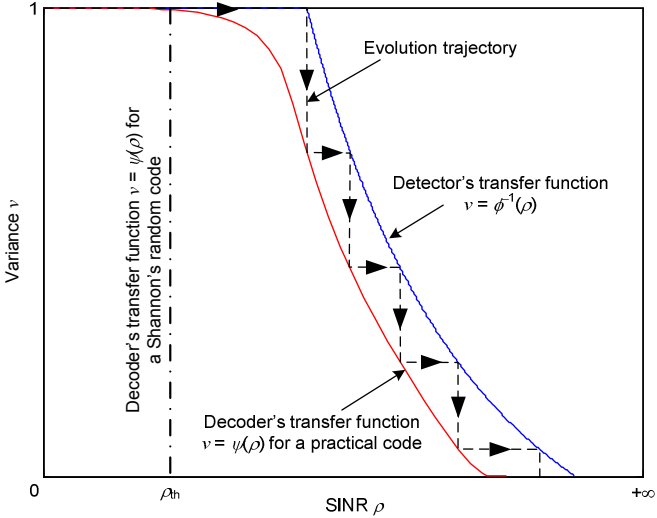

Fig. 2. An illustration of the SINR-variance transfer chart.

## IV. AREA THEOREMS AND PRECODER OPTIMIZATION

In this section, we establish area theorems for the proposed LP-LMMSE scheme. We further discuss the optimization of the power matrix $W$.

### A. Area Property

The main theorem of this paper is presented below.

*Theorem 1:* As $J \to \infty$, if the detector and the decoder are matched, then an achievable rate (per antenna per channel use) of the LP-LMMSE scheme is given by

$$R = \frac{1}{N}\sum_{n=1}^{N}\log\left(1+\frac{d_n^2}{\sigma^2}\right). \quad (25)$$

The proof of Theorem 1 is given in Appendix C. Note that (25) is the input-output mutual information of the channel in (13b) with the entries of $x$ i.i.d. drawn from $\mathcal{CN}(0, 1)$.

Theorem 1 is based on Assumptions 1 and 2. As discussed in Section II, these assumptions can be approximately ensured by adopting sufficiently long interleaving.

### B. Water-Filling Precoding

Now we consider the optimization of the power matrix $W$. Recall from (13a) that $D = \Lambda W^{1/2}$, or equivalently, $d_i = \lambda_i w_i^{1/2}$, for $i = 1, …, N$. Then, (25) becomes

$$R = \frac{1}{N}\sum_{n=1}^{N}\log\left(1+\frac{\lambda_n^2 w_n}{\sigma^2}\right). \quad (26)$$

We aim at maximizing the above rate over $W$ subject to the power constraint in (11). Clearly, the solution coincides with the renowned water-filling power allocation [44].

*Corollary 1:* Under the conditions of Theorem 1, if $W$ follows water-filling power allocation, then the LP-LMMSE scheme achieves the water-filling capacity of the channel in (8a).

Corollary 1 shows that linear precoding, together with proper error-control coding and iterative LMMSE detection, can potentially achieve the water-filling channel capacity. This achievability is based on a single code with a matched decoding transfer function.

### C. Area Theorem for Constrained Signaling

So far, we have shown that the proposed linear precoding scheme is asymptotically optimal (in the sense of achieving the water-filling capacity) as the signaling approaches Gaussian. However, in practice, the signaling is usually constrained on a finite-size discrete constellation. We next establish an area theorem for constrained signaling.

Define the $\gamma$-function as

$$\gamma(\rho) = \mathrm{E}\left[\left|x - \mathrm{E}[x \mid x+\eta]\right|^2\right] \quad (27)$$

where $x$ is uniformly taken over $\mathcal{S}$, and $\eta$ is independently drawn from $\mathcal{CN}(0, 1/\rho)$. We have the following result with the proof given in Appendix D.

*Theorem 2:* Assume that the detector's inputs $\{a_i\}$ are modeled as observations of $\{x_i\}$ from an effective AWGN channel. Then, as $J \to \infty$, if the detector and the decoder are matched, an achievable rate of the LP-LMMSE scheme is

$$R = \log|\mathcal{S}| - \int_0^{+\infty}\gamma(\rho+\phi(\gamma(\rho)))\mathrm{d}\rho \quad (28)$$

where the $\phi$-function is given by (15), and $\gamma$ is defined in (27).

Theorem 2 holds under the AWGN assumption on $\{a_i\}$. It was observed that this assumption is empirically true for BPSK, QPSK, and SCM (with BPSK/QPSK layers). However, this assumption may be far from true if other modulation techniques, such as bit-interleaved coded modulation [47], are employed. In the later cases, the area theorem based on the measure of mean-square error (MSE) established in [48] can be used for performance evaluation. Nevertheless, the related discussion is out of the scope of this paper.

### D. Precoder Optimization for Constrained Signaling

Recall that, for Gaussian signaling, the optimal power matrix $W$ is the water-filling solution. However, water-filling is not necessarily optimal for discrete signaling. Based on the area property in Theorem 2, we can optimize $W$ to maximize the achievable rate of the LP-LMMSE scheme. This problem can be formulated as:

maximize $\quad \log|\mathcal{S}| - \int_0^{+\infty}\gamma(\rho+\phi(\gamma(\rho)))\mathrm{d}\rho \quad (29\mathrm{a})$

subject to $\quad N^{-1}\sum_{n=1}^{N} w_n \leq P. \quad (29\mathrm{b})$

In the above, $P$ is the maximum signal power given in (8c). The following result is useful in solving (29).

*Lemma 3:* The rate $R$ in (28) is a concave function of $\{w_1, …, w_N\}$, provided that $\gamma(\rho)$ is a convex function of $\rho$.

The proof of Lemma 3 can be found in Appendix E. The $\gamma$-function is convex for most commonly used signaling constellations [43]. From Lemma 3, the problem in (29) can be solved using standard convex programming [49].

## V. EXTENSIONS TO MIMO CHANNELS WITH CSIT UNCERTAINTY

In this section, we extend the results in Sections III and IV to more general situations that CSI is not perfectly known at the transmitter.

### A. Linear Precoding with Imperfect CSIT

We now consider MIMO systems with partial CSIT. Here, partial CSIT means that the channel matrices are not exactly known, instead, only the statistics of the channel is known at the transmitter side. The precoding technique in Section III.B requires perfect CSIT, and so cannot be directly applied here. The following is a modified solution.

Return to the extended system in (8a)

$$\tilde{y} = \tilde{H}\tilde{x} + \tilde{\eta}, \tag{30a}$$

where the extended channel contains $J/N$ channel realizations:

$$\tilde{H} = \begin{bmatrix} H_1 & & & \\ & H_2 & & \\ & & \ddots & \\ & & & H_{J/N} \end{bmatrix}. \tag{30b}$$

We focus on the ergodic case, in which $\{H_i\}$ are independent realizations of an $M$-by-$N$ random matrix, with abuse of notation, denoted by $H$. This model also includes orthogonal frequency-division multiplexing (OFDM) systems [40] with independent fading over different sub-carriers.

The channel input $\tilde{x}$ is related to $x$ as

$$\tilde{x} = \tilde{P}\tilde{\Pi}\tilde{F}x \tag{31a}$$

with

$$\tilde{P} = \text{diag}\{P, P, \ldots, P\} \tag{31b}$$

$$\tilde{F} = \text{diag}\{F, F, \ldots, F\}. \tag{31c}$$

where $\tilde{P}$ is a $J$-by-$J$ block-diagonal matrix with $P$ of size $N$-by-$N$, $\tilde{\Pi}$ is a $J$-by-$J$ permutation matrix, and $\tilde{F}$ is a $J$-by-$J$ block-diagonal matrix with each block $F$ being the normalized DFT matrix of size $L$-by-$L$. Combining (30a) and (31) and letting $y = \tilde{y}$ and $\eta = \tilde{\eta}$, we obtain an equivalent channel as

$$y = Ax + \eta \tag{32a}$$

with

$$A = \tilde{H}\tilde{P}\tilde{\Pi}\tilde{F}. \tag{32b}$$

The power constraint now becomes

$$N^{-1}\text{tr}\{PP^H\} \leq P. \tag{33}$$

The precoding matrix $\tilde{P} = \text{diag}\{P, \ldots, P\}$ allows the precoder to exploit the benefit provided by the available CSIT.[4] The optimization of $P$ is briefly discussed in Subsection C.

The precoding matrix $\tilde{F}$ ensures that every coded symbol in $x$ is sufficiently dispersive over time and space. Thus, it is required that the DFT size $L$ is sufficiently large. A convenient choice of $L$ is $L = J/N$, and then there are $N$ length-$L$ DFT matrices in $\tilde{F}$.

Now we consider the design of the permutation matrix $\tilde{\Pi}$. Let $q = [q_1^T, \ldots, q_N^T]^T = \tilde{F}x$ and $x = [x_1^T, \ldots, x_N^T]^T$, where each DFT block $q_i = Fx_i$ is of size $L$-by-1. Also denote $c = [c_1^T, \ldots, c_{J/N}^T]^T = \tilde{\Pi}q$, where each $c_i$ is an $N$-by-1 vector. The two criteria for $\tilde{\Pi}$ are listed below.

(i) For each $i$, the entries of $q_i$ are transmitted at different channel uses, i.e., no two entries of $q_i$ are connected to a same $c_j$ for any index $j = 1, \ldots, J/N$; and

(ii) Treat $PH_k$ as the $k$th realization of the equivalent channel $PH$. Then, any $N$ consecutive entries of $q_i$ are transmitted at different *transmit antennas* of the equivalent channel $PH$. As a result, for each index $j$, the set of {the $j$th entry of $c_k | k = 1, \ldots, J/N$} contains $J/N^2$ entries of $q_i$, $i = 1, \ldots, N$.[5]

The choice of $\tilde{\Pi}$ satisfying the above criteria is not unique. A simple choice is as follows: for each index $k$, the $k$th entry of $q_1$ is connected to the $(k \mod N)$th input of the equivalent channel $PH_k$; then, for $j = 2, \ldots, J/N$, the connection pattern of each $q_j$ is just a one-entry cyclic shift of the previous one. An example for $J = 16$ and $N = 4$ is illustrated in Fig. 3.

The above choice of $\tilde{F}$ and $\tilde{\Pi}$ ensures that the behavior of the iterative receiver can be characterized by a single-variable recursion, as will be detailed in the next subsection.

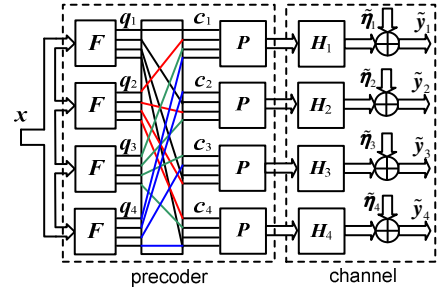

Fig. 3. An illustration of the proposed precoder for imperfect CSIT with $J = 16$ and $N = 4$.

### B. SINR-Variance Transfer Chart

Iterative LMMSE detection described in Section II is applied to the system in (32). We next show that the outputs of the detector can still be characterized by a single SINR.

With abuse of notation, define

$$\phi(v) = \left(\frac{1}{N}\mathbb{E}\left[\text{tr}\left\{\left(\frac{1}{v}I + \frac{1}{\sigma^2}HQH^H\right)^{-1}\right\}\right]\right)^{-1} - \frac{1}{v}. \tag{34}$$

where $Q = PP^H$, and the expectation is taken over the distribution of $H$. We aim to show that

$$\rho_i = 1/u_i = \phi(v), \text{ for } i = 1, \ldots, J. \tag{35}$$

With (32b), we can rewrite (5c) as

$$M = \tilde{F}^H\tilde{\Pi}^H\left(vI - v^2\tilde{P}^H\tilde{H}^H(v\tilde{H}\tilde{P}\tilde{P}^H\tilde{H}^H + \sigma^2 I)^{-1}\tilde{H}\tilde{P}\right)\tilde{\Pi}\tilde{F}$$

$$= \tilde{F}^H B\tilde{F}$$

where

$$B = \tilde{\Pi}^H\left(vI - v^2\tilde{P}^H\tilde{H}^H(v\tilde{H}\tilde{P}\tilde{P}^H\tilde{H}^H + \sigma^2 I)^{-1}\tilde{H}\tilde{P}\right)\tilde{\Pi}. \tag{36}$$

We can express $M$ and $B$ in a block-wise form with each block of size $(J/N)$-by-$(J/N)$. Let $M_i$ and $B_i$ be the $i$th diagonal block of $M$ and $B$, respectively. Then, we obtain

$$(M)_{\text{diag}} = \begin{bmatrix} (F^H B_1 F)_{\text{diag}} & & \\ & \ddots & \\ & & (F^H B_n F)_{\text{diag}} \end{bmatrix},$$

where $(\cdot)_{\text{diag}}$ returns a diagonal matrix specified by the diagonal of the matrix in the parenthesis. Recall that both $\tilde{H}$ and $\tilde{P}$ are block diagonal, and so is $\tilde{P}^H \tilde{H}^H (v \tilde{H} \tilde{P} \tilde{P}^H \tilde{H}^H + \sigma^2 I)^{-1} \tilde{H} \tilde{P}$. Thus, from the criterion (i) of $\tilde{\Pi}$ in Subsection A, every $B_i$ is diagonal. Therefore, as $J \to \infty$

$$M_i(k,k) \stackrel{(a)}{=} \frac{N}{J} \text{tr}\{B_i\} \stackrel{(b)}{\to} \frac{1}{J} \text{tr}\{B\}$$

$$\stackrel{(c)}{=} \frac{1}{J} \text{tr}\{vI - v^2 \tilde{P}^H \tilde{H}^H (v \tilde{H} \tilde{P} \tilde{P}^H \tilde{H}^H + \sigma^2 I)^{-1} \tilde{H} \tilde{P}\}$$

$$= \frac{1}{J} \text{tr}\left\{\left(\frac{1}{v} I + \frac{1}{\sigma^2} \tilde{H} \tilde{P} \tilde{P}^H \tilde{H}^H\right)^{-1}\right\}$$

$$\stackrel{(d)}{=} \frac{1}{J} \sum_{i=1}^{J/N} \text{tr}\left\{\left(\frac{1}{v} I + \frac{1}{\sigma^2} H_i P P^H H_i^H\right)^{-1}\right\}$$

$$\to \frac{1}{N} \text{E}\left[\text{tr}\left\{\left(\frac{1}{v} I + \frac{1}{\sigma^2} HQH^H\right)^{-1}\right\}\right] \quad (37)$$

where step (a) follows from the fact that, for each $i$

$$(F^H B_i F)_{\text{diag}} = \left(\frac{N}{J} \sum_{n=1}^{N} B_i(n,n)\right) I = \frac{N}{J} \text{tr}\{B_i\} I,$$

step (b) from the criterion (ii) of $\tilde{\Pi}$, step (c) from (36), and step (d) by substituting $\tilde{H} = \text{diag}\{H_1, \ldots, H_{J/N}\}$, $\tilde{P} = \text{diag}\{P, \ldots, P\}$, and $Q = PP^H$. Substituting (37) into (5a), we arrive at (35).

---

[4] We emphasize that $P$ is designed to be adaptive to the available CSI. Here, $P$ remains constant for different channel uses, as the channel statistics does not change. However, if the channel statistics varies, $P$ should vary accordingly.

[5] We always assume that $J$ is properly chosen such that $J/N^2$ is an integer.

Now, similar to Lemma 1, we have the following result.

*Lemma 4:* The detector's output $b_i$ can be expressed as $b_i = x_i + n_i$, where $n_i$ is independent of $x_i$ and converges to $\mathcal{CN}(0, 1/\phi(v))$ as $J \to \infty$.

The above lemma is literally the same as Lemma 1, except that $\phi(v)$ here is given by (34). The proof mostly follows that of Lemma 1. We omit the details for simplicity. With Lemma 4, we can still characterize the behavior of the iterative receiver using the SINR-variance transfer functions, namely, $\rho = \phi(v)$ and $v = \psi(\rho)$, as described in Section III.E.

### C. Area Theorem and Precoder Optimization

Now we are ready to present the following result.

*Theorem 3:* As $J \to \infty$, if the detector and the decoder are matched, an achievable information rate (per transmit antenna per channel use) of the LP-LMMSE scheme is given by

$$R = \frac{1}{N} \text{E}\left[\log \det\left(I + \frac{1}{\sigma^2} HQH^H\right)\right]. \quad (38)$$

The proof of Theorem 3 is given in Appendix F.

With Theorem 3, we can formulate the following rate-maximization problem:

maximize $\quad \text{E}\left[\log \det\left(I + \frac{1}{\sigma^2} HQH^H\right)\right]$ (39a)

subject to $\quad N^{-1} \text{tr}\{Q\} \leq P$. (39b)

The above optimization problem is convex, and thus can be solved numerically using standard convex programming tools. Moreover, the explicit solutions to (39) in a variety of CSIT settings have been studied in the literature. We refer interested readers to [34] [50] [51] for more details.

For constrained signaling, it is straightforward to show that Theorem 2 holds literally for the partial CSIT case, except that $\phi(v)$ is replaced by (34). Then, we can formulate an optimization problem similar to (29), which is solvable using standard convex programming. Details are omitted for simplicity.

## VI. NUMERICAL RESULTS

In this section, we provide numerical examples to demonstrate the achievable rates of the proposed scheme. Note that the channel SNR is defined as

$$\text{SNR} = PN/\sigma^2.$$

We first consider the case of full CSIT in a randomly generated 2×2 MIMO 3-tap ISI channel with the tap coefficients

$$\begin{bmatrix} 0.5339 + j0.5395 & -0.4245 + j0.0648 \\ -0.3347 - j0.3727 & -0.4672 - j0.2420 \end{bmatrix}, \quad (40a)$$

$$\begin{bmatrix} 0.0582 - j0.2706 & 0.1525 - j0.7565 \\ -0.4968 - j0.1543 & -0.5243 - j0.5915 \end{bmatrix}, \quad (40b)$$

and $\begin{bmatrix} -0.5262 - j0.2654 & -0.3714 - j0.2865 \\ 0.6721 - j0.1635 & 0.1607 - j0.2695 \end{bmatrix}. \quad (40c)$

The DFT is applied to convert the above MIMO ISI channel to a set of $J = 256$ 2-by-2 parallel MIMO channels. Note that the effect of cyclic prefix is ignored here.

Fig.4 shows the achievable rates of the scheme with various precoders. The flat, water-filling, and optimized precoders are considered. For flat precoding, $W = \varepsilon I$ where $\varepsilon$ is a scaling factor to meet the power budget; for water-filling precoding, $W$ is given by the water-filling solution; and for optimized precoding, $W$ is obtained by solving (29). QPSK modulation and standard 16-QAM (cf., SCM-1 in [57]) are employed for signaling.

From Fig.4, the optimized and water-filling precoders have similar performance, and both considerably outperform the flat precoder in the low SNR region. This implies that the water-filling precoder provides an attractive low-complexity near-optimum option for discrete signaling. Fig.4 also includes the Gaussian water-filling capacity as a reference. We see that the optimized and water-filling precoders approach the water-filling capacity in the low SNR region.

Fig.4 also demonstrates that the achievable rate of the LP-LMMSE scheme is significantly increased by changing the signaling constellation from QPSK to 16-QAM. We see that the gap between the achievable rate of LP-LMMSE with 16-QAM and the water-filling capacity is not significant for a rate up to 4 bits. For a higher transmission rate, a larger signaling constellation is necessary.

We now provide code design examples to verify that the theoretical limits given by our analysis are indeed approachable. We basically follow the design approach described in [45]. Here, we only provide the design results. We consider standard 16-QAM signaling with 2-layer SCM. Each SCM layer is QPSK modulated with Gray mapping. (Note: for standard 16-QAM signaling, the power ratio of the two SCM layers is 4:1.) Two irregular LDPC codes are designed based on the curve-matching principle. For the SCM layer with higher power, the edge degree distributions of the LDPC code are given by $\{\lambda_1(x) = 0.3495x+0.2142x^2+0.1165x^6+0.0857x^7+0.1331x^{17} +0.1010x^{20}, \rho_1(x) = x^6\}$; for the other layer, the edge degree

distributions of the LDPC code are $\{\lambda_2(x) = 0.3907x +0.1854x^2+0.0968x^{10}+0.1877x^{11}+0.1394x^{33}, \rho_2(x) = x^6\}$. The system throughput is 4 bits per channel use. The transfer curves of the detector and the decoder are illustrated in Fig.5, and the simulated system performance is given in Fig.6. From Fig.6, the design threshold is 6.7 dB, 1.1 dB away from the water-filling capacity, 0.5 dB away from the performance limit of the LP-LMMSE scheme with standard 16-QAM. The simulated system performance with code length $\approx 10^5$ at BER = $10^{-4}$ is about 7.0 dB, only 0.3 dB away from the design threshold.

It is interesting to pay special attention to point *A* in Fig.5. This is the transition point where, among two layers of the LDPC codes, the one with higher power is nearly fully decoded (i.e., with very reliable outputs) and the one with lower power starts to be decoded.

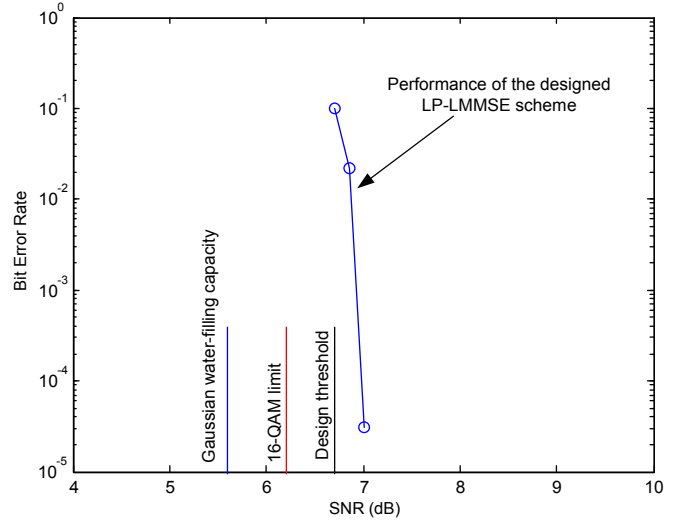

Fig. 6. The BER performance of the LP-LMMSE scheme with 2-layer SCM over the MIMO ISI channel in (40). Water-filling precoding and standard 16-QAM constellation are used. Some simulation settings are $M = N = 2$, $J = 256$, and $K = 256$.

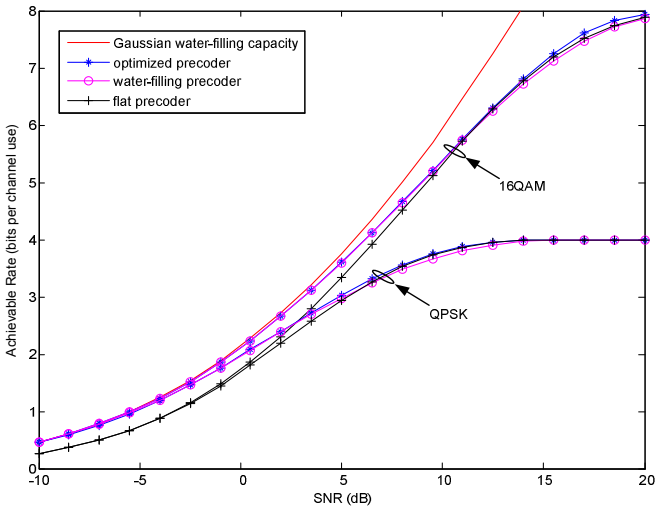

Fig. 4. The achievable rates of the proposed scheme in the MIMO-ISI channel in (40). The flat, optimized, and water-filling precoders are considered. The water-filling capacity and QPSK i.i.d. capacity are also included for references.

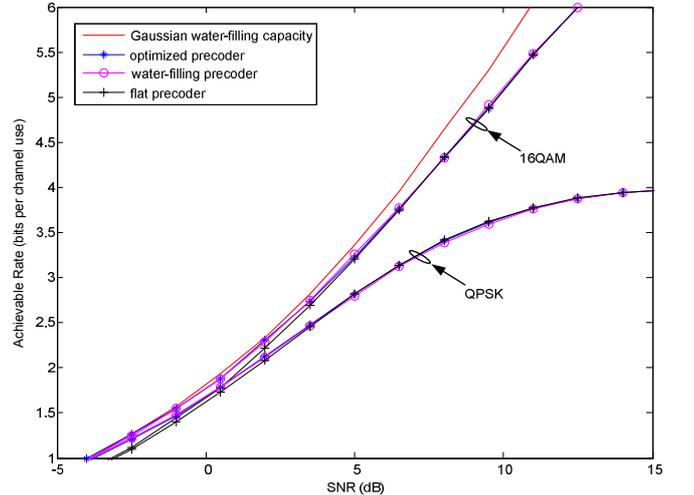

Fig. 7. The achievable rates of the LP-LMMSE scheme in the flat-fading 2-by-2 MIMO channel with partial CSIT (of $\theta = 0.5$). The flat, optimized, and water-filling precoders are considered. The unconstrained channel capacity is also included as a reference.

Now consider the case of partial CSIT. We assume the mean-feedback model [53], i.e., each $H_i$ is a random realization of
$$H = \mathrm{E}[H] + \Delta H$$
where the entries of $\mathrm{E}[H]$ are independently drawn from $\mathcal{CN}(0,\theta)$ with $\theta \in [0, 1]$, and those of $\Delta H$ from $\mathcal{CN}(0, 1-\theta)$. In simulation, $\theta$ is set to 0.5. The channel mean $\mathrm{E}[H]$ remains constant for each frame, but varies independently from frame to frame. Each frame contains 50 independently generated $\Delta H$. The achievable rates of the LP-LMMSE scheme (averaged over 2000 frames) are illustrated in Fig.7. We see that similar trends as in Fig. 4 have been observed.

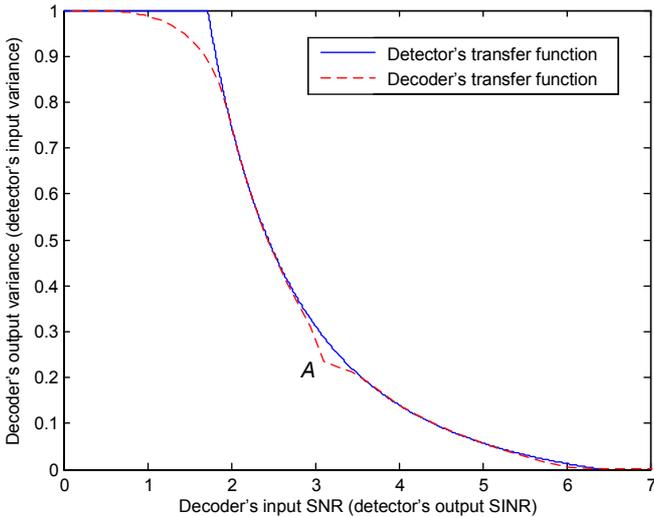

Fig. 5. The SNR-variance transfer functions of the detector and the decoder for the LP-LMMSE scheme with 2-layer SCM over the MIMO ISI channel in (40). Water-filling precoding and standard 16-QAM constellation are used. The transfer function of the detector is given by (15) at the channel SNR = 6.7 dB. The transfer function of the decoder is obtained by simulation.

## VII. CONCLUSIONS

In this paper, we establish analytical techniques to evaluate the achievable rate of iterative MIMO systems with multi-ary signaling. The main results are summarized as follows.

We first discuss the case of full CSIT. We propose a linear precoding technique and show that the corresponding LP-LMMSE scheme can be analyzed using the SINR-variance transfer chart. An area theorem is then established, which reveals that the proposed LP-LMMSE scheme is information lossless for unconstrained signaling. We also develop an area theorem to evaluate the achievable rate of the LP-LMMSE scheme with finite discrete signaling. Based on the established area theorems, the precoder optimization is discussed. It is shown that a properly designed LP-LMMSE scheme can achieve the water-filling channel capacity.

We then consider the situation of partial CSIT. We show that, with the modified linear precoding technique, our previous results for the case of full CSIT can be extended to the case of partial CSIT. Particularly, we show that a properly designed LP-LMMSE scheme can achieve the ergodic capacity of the fading MIMO channel.

## APPENDIX A
### PROOF OF LEMMA 1

Substituting (4b) and $\boldsymbol{A} = \boldsymbol{D}\boldsymbol{F}$ into (4a), we obtain the $i$th entry of $\hat{\boldsymbol{x}}$ as

$$\begin{aligned}\hat{x}_i &= \bar{x}_i + v\boldsymbol{f}_i^H \boldsymbol{D}^H (v\boldsymbol{D}\boldsymbol{D}^H + \sigma^2 \boldsymbol{I})^{-1}(\boldsymbol{y} - \boldsymbol{A}\bar{\boldsymbol{x}}) \\ &= \left(1 - v\boldsymbol{f}_i^H \boldsymbol{D}^H (v\boldsymbol{D}\boldsymbol{D}^H + \sigma^2 \boldsymbol{I})^{-1} \boldsymbol{D}\boldsymbol{f}_i\right)\bar{x}_i \\ &\quad + v\boldsymbol{f}_i^H \boldsymbol{D}^H (v\boldsymbol{D}\boldsymbol{D}^H + \sigma^2 \boldsymbol{I})^{-1}(\boldsymbol{y} - \boldsymbol{A}\bar{\boldsymbol{x}}_{\setminus i}) \\ &= v^{-1} M(i,i)\bar{x}_i + v\boldsymbol{f}_i^H \boldsymbol{D}^H (v\boldsymbol{D}\boldsymbol{D}^H + \sigma^2 \boldsymbol{I})^{-1}(\boldsymbol{y} - \boldsymbol{A}\bar{\boldsymbol{x}}_{\setminus i})\end{aligned} \quad (A1)$$

where $\bar{\boldsymbol{x}}_{\setminus i}$ represents the vector obtained by setting the $i$th entry of $\bar{\boldsymbol{x}}$ to zero, and the last equality follows from (14b). Then

$$\begin{aligned}b_i &\stackrel{(a)}{=} \frac{v}{M(i,i)\rho} \boldsymbol{f}_i^H \boldsymbol{D}^H (v\boldsymbol{D}\boldsymbol{D}^H + \sigma^2 \boldsymbol{I})^{-1}(\boldsymbol{y} - \boldsymbol{A}\bar{\boldsymbol{x}}_{\setminus i}) \\ &\stackrel{(b)}{=} \frac{v}{M(i,i)\rho} \boldsymbol{f}_i^H \boldsymbol{D}^H (v\boldsymbol{D}\boldsymbol{D}^H + \sigma^2 \boldsymbol{I})^{-1} \boldsymbol{D}\boldsymbol{f}_i x_i + n_i \\ &\stackrel{(c)}{=} x_i + n_i\end{aligned}$$

where step (a) follows by substituting (A1) into (5b) and by noting $u_i = 1/\rho$, (b) by substituting (13b) and (16), and (c) from (14a) and (15). Moreover, it can be verified that

$$\mathrm{E}[n_i] = 0 \text{ and } \mathrm{E}[|n_i|^2] = \phi(v)^{-1}.$$

We next show that, for a sufficiently large $J$, the distribution of $n_i$ converges to a normal distribution. This basically follows the central limit theorem. The details are as follows.

Recall the expression of $n_i$ in (16). As $\boldsymbol{\eta}$ is Gaussian, we only need to show that

$$e_i = \boldsymbol{f}_i^H \boldsymbol{D}(v\boldsymbol{D}^H\boldsymbol{D} + \sigma^2 \boldsymbol{I})^{-1} \boldsymbol{D}^H \boldsymbol{F}(\boldsymbol{x}_{\setminus i} - \bar{\boldsymbol{x}}_{\setminus i})$$

converges to a normal distribution. Let $\boldsymbol{s}$ be the diagonal of $\boldsymbol{D}^H (v\boldsymbol{D}\boldsymbol{D}^H + \sigma^2 \boldsymbol{I})^{-1} \boldsymbol{D}$, and let

$$\boldsymbol{t} = J^{-1/2} \boldsymbol{F}^H \boldsymbol{s}. \quad (A2)$$

It is straightforward to verify that

$$t_k = J^{-1/2} \boldsymbol{f}_k^H \boldsymbol{s} = \boldsymbol{f}_i^H \boldsymbol{D}(v\boldsymbol{D}^H\boldsymbol{D} + \sigma^2 \boldsymbol{I})^{-1} \boldsymbol{D}^H \boldsymbol{f}_{i+k-1}.$$

With the above, $e_i$ can be equivalently expressed as

$$e_i = \sum_{k=2}^{J} t_k \left(x_{[i+k-1]} - \bar{x}_{[i+k-1]}\right) \quad (A3)$$

where $x_{[i+k-1]} = x_{i+k-1-J}$ if $i+k-1 > J$.

Both $\boldsymbol{x}$ and $\bar{\boldsymbol{x}}$ are i.i.d. sequences, and so is $\boldsymbol{x} - \bar{\boldsymbol{x}}$. We consider the following limit:

$$\begin{aligned}\omega_\varepsilon &= \lim_{J \to \infty} \frac{\sum_{i=2}^{J} \mathrm{E}\left|t_i\left(x_{[i+k-1]} - \bar{x}_{[i+k-1]}\right)\right|^{2+\varepsilon}}{\left(\sum_{i=2}^{J} \mathrm{E}\left|t_i\left(x_{[i+k-1]} - \bar{x}_{[i+k-1]}\right)\right|^2\right)^{\frac{2+\varepsilon}{2}}} \\ &= \frac{\mathrm{E}|x_1 - \bar{x}_1|^{2+\varepsilon}}{\left(\mathrm{E}|x_1 - \bar{x}_1|^2\right)^{\frac{2+\varepsilon}{2}}} \cdot \lim_{J \to \infty} \frac{\sum_{k=2}^{J}|t_k|^{2+\varepsilon}}{\left(\sum_{k=2}^{J}|t_k|^2\right)^{\frac{2+\varepsilon}{2}}}\end{aligned} \quad (A4)$$

where $\varepsilon$ is a positive number, and the second equality follows from the fact that

$$\mathrm{E}\left[|x_i - \bar{x}_i|^{2+\varepsilon}\right] = \mathrm{E}\left[|x_i - \mathrm{E}[x_i \mid a_i]|^{2+\varepsilon}\right] = c_\varepsilon, \text{ for any } i,$$

with $c_\varepsilon$ invariant to the index $i$. From the Lyapunov's central limit theorem, if $\omega_\varepsilon = 0$ for some $\varepsilon > 0$, $e_i$ in (A3) converges in distribution to a normal distribution (p. 309, [54]).

What remains is to verify $\omega_\varepsilon = 0$ for $\varepsilon = 2$. From (A4), it suffices to show that

$$\lim_{J \to \infty} \frac{\sum_{k=2}^{J}|t_k|^4}{\left(\sum_{k=2}^{J}|t_k|^2\right)^2} = 0. \quad (A5)$$

To this end, define $\Delta \boldsymbol{s} = \boldsymbol{s} - \bar{s}\boldsymbol{1}$, where $\boldsymbol{s}$ is the diagonal of $\boldsymbol{D}^H(v\boldsymbol{D}\boldsymbol{D}^H + \sigma^2 \boldsymbol{I})^{-1}\boldsymbol{D}$, $\bar{s}$ is the average value of the entries of $\boldsymbol{s}$, and $\boldsymbol{1}$ represents an all-one vector. From the discussions below (9), the ordering of the diagonal entries of $\boldsymbol{\Lambda}$ are chosen such that the entries of $\Delta \boldsymbol{s}$ are asymptotically uncorrelated, i.e., for any $i' \neq 0$,

$$\lim_{J \to \infty} \frac{\sum_{i=1}^{J} \Delta s_i \Delta s_{i+i'}}{\sum_{i=1}^{J} \Delta s_i^2} = 0. \quad (A6)$$

From (A2), we can express

$$t_k = \frac{1}{J} \sum_{i=1}^{J} \Delta s_i \exp(j2\pi(i-1)(k-1)/J), \text{ for } k = 2, \ldots, J.$$

Then, with some straightforward but tedious derivations, we can obtain (A5), which completes the proof of Lemma 1. ∎

## APPENDIX B
### PROOF OF LEMMA 2

We prove Lemma 2 by constructing a series of $n$-layer SCM code $\Gamma_n$ as follows. Let $\rho_{\max}$ be an arbitrary positive number. We quantize the SNR range $(0, \rho_{\max}]$ with interval $\Delta r \equiv \rho_{\max}/n$ as: $r_0 = 0, r_1 = \rho_{\max}/n, \ldots, r_n = \rho_{\max}$. Define

$$p_i = \psi(r_i), \text{ for } i = 0, 1, \ldots, n. \quad (A7)$$

Note that $\Delta p_i = p_{i-1} - p_i \geq 0$ since $\psi(\rho)$ is monotonically decreasing. Let $\Delta p_i$ be the signal power of the $i$th layer of $\Gamma_n$, for $i = 1, \ldots, n$. The $i$th layer of $\Gamma_n$ is encoded using a random codebook drawn from $\mathcal{CN}(0, \Delta p_i)$ at a rate of

$$R_{n,i} = \log\left(1 + \frac{\Delta p_i}{p_{i+1} + r_i^{-1}}\right), \text{ for } i = 1, \ldots, n. \quad (A8)$$

Here, for simplicity of discussion, we assume that each SCM layer employs Gaussian signaling. The extension to discrete signaling will be discussed at the end of the proof. We have the following result.

*Fact 1:* For any $i \in \{0, 1, \ldots, n\}$ and the decoder's input SNR $\rho \in [r_i, r_{i+1})$, the first $i$ layers of $\Gamma_n$ can be successively decoded in the order from 1 to $i$.

*Proof of Fact 1:* We prove by induction. Fact 1 trivially holds for $i = 0$. Now suppose that Fact 1 holds for $i = k-1$. Consider any input SNR $\rho \in [r_k, r_{k+1})$. By assumption, the first $k-1$ layers are decodable. After decoding and canceling the first $k-1$ layers from the decoder's input, the SINR seen by the $k$th layer is

$$\frac{\Delta p_k}{p_{k+1} + \rho^{-1}} > \frac{\Delta p_k}{p_{k+1} + r_k^{-1}}, \quad (A9)$$

where $\rho^{-1}$ is the noise power (as the decoder's input SNR is $\rho$ and the signal power is normalized to 1), and the inequality follows from $\rho \in (r_k, r_{k+1}]$. Thus, from the well-known Shannon's capacity formula, the $k$th layer with rate $R_{n,i}$ in (A8) is decodable, hence the proof of Fact 1.

Now consider the SINR-variance transfer function of $\Gamma_n$, denoted by $\psi_n(\rho)$. The SINR-variance transfer function of a random code with a sufficiently large code length is given by (cf., [14, Theorem 3])

$$\psi_{\text{random}}(\rho) = \begin{cases} p, & \text{for } \rho < \rho_{\text{th}} \\ 0, & \text{for } \rho > \rho_{\text{th}} \end{cases}$$

where $p$ is the transmission power and $\rho_{\text{th}}$ is the SNR threshold stipulated by the channel capacity (cf., Fig. 2). Thus, together with Fact 1, it can be readily verified that $\psi_n(\rho)$ has a ladder-shaped form (as illustrated in Fig. 8) satisfying

$$\psi_n(\rho) \leq \psi(\rho), \text{ for any } \rho > 0.$$

What remains is to show that $R_n \to R$ as $n \to \infty$. To this end, we have

$$\lim_{n\to\infty} R_n = \lim_{n\to\infty} \sum_{i=1}^n R_{n,i} \stackrel{(a)}{=} \lim_{n\to\infty} \sum_{i=1}^n \left(\frac{\Delta p_i}{p_{i+1} + r_i^{-1}} + o\left(\frac{\Delta p_i}{p_{i+1} + r_i^{-1}}\right)\right)$$

$$\stackrel{(b)}{=} \lim_{n\to\infty} \sum_{i=1}^n \frac{\psi(r_{i-1}) - \psi(r_i)}{\psi(r_{i+1}) + r_i^{-1}} \stackrel{(c)}{=} -\lim_{n\to\infty} \sum_{i=1}^n \frac{\psi'(r_i)(r_i - r_{i-1})}{\psi(r_{i+1}) + r_i^{-1}}$$

$$= -\int_0^{\rho_{\max}} \frac{\rho \psi'(\rho)}{\rho \psi(\rho) + 1} d\rho = \int_0^{\rho_{\max}} \frac{\psi(\rho)}{\rho \psi(\rho) + 1} d\rho - \int_0^{\rho_{\max}} \frac{\rho \psi'(\rho) + \psi(\rho)}{\rho \psi(\rho) + 1} d\rho$$

$$= \int_0^{\rho_{\max}} \frac{\psi(\rho)}{\rho \psi(\rho) + 1} d\rho - \log(\rho \psi(\rho) + 1)\big|_{\rho=0}^{\rho=\rho_{\max}}$$

$$= \int_0^{\rho_{\max}} \frac{1}{\psi(\rho)^{-1} + \rho} d\rho - \log(\rho_{\max} \psi(\rho_{\max}) + 1) \quad (A10)$$

where step (*a*) follows from (A8), step (b) utilizes (A7), and step (*c*) utilizes the fact that $\psi(\rho)$ is first-order differentiable.

Recall that $\rho_{\max}$ is arbitrary. Letting $\rho_{\max} \to \infty$, together with (22), we obtain from (A10) that

$$\lim_{n\to\infty} R_n = \int_0^\infty (\psi(\rho)^{-1} + \rho) d\rho.$$

So far, we have shown that Lemma 2 holds when the SCM layers employ Gaussian signaling. The proof for discrete signaling follows from the fact that, when the signal power tends to zero, the mutual information on a Gaussian channel is not sensitive to the input distribution. Specifically, from [15, Lemma 1], for any input distribution with zero mean, the capacity of the AWGN channel with SNR = $t$ is given by $t+o(t)$. Then, for discrete signaling, we only need to replace (A8) by

$$R_{n,i} = \frac{\Delta p_i}{p_{i+1} + r_i^{-1}} + o\left(\frac{\Delta p_i}{p_{i+1} + r_i^{-1}}\right), \text{ for } i = 1, \ldots, n.$$

This completes the proof of Lemma 2. ∎

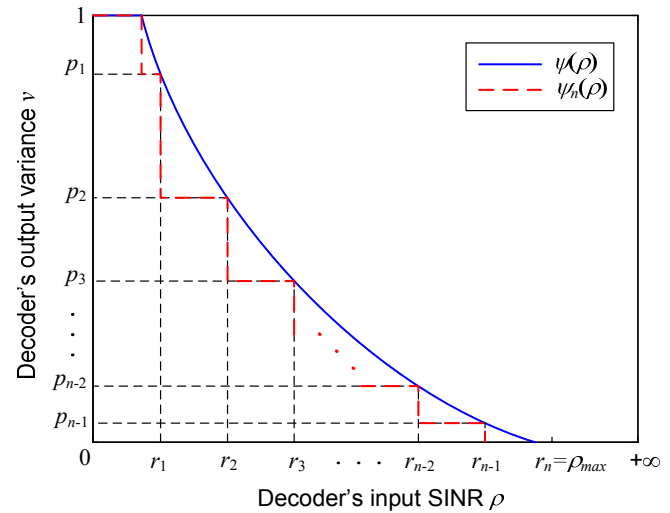

Fig. 8. An illustration of the target SINR-variance transfer curve $\psi(\rho)$ and the corresponding transfer curve $\psi_n(\rho)$ of the matched $n$-layer SCM code.

## APPENDIX C
### PROOF OF THEOREM 1

It can be verified that $\psi(\rho) = \phi^{-1}(\rho)$ satisfies the regularity conditions required in Lemma 2. Thus, an achievable rate of the LP-LMMSE scheme is given by

$$R \stackrel{(a)}{=} \int_0^{+\infty} \left((\psi(\rho))^{-1} + \rho\right)^{-1} d\rho$$

$$\stackrel{(b)}{=} \int_{\phi(1)}^{\phi(0)} \frac{1}{(\phi^{-1}(\rho))^{-1} + \rho} d\rho + \int_0^{\phi(1)} \frac{1}{1+\rho} d\rho$$

$$\stackrel{(c)}{=} \int_{v=1}^{v=0} \frac{1}{v^{-1} + \phi(v)} d\phi(v) + \log(1 + \phi(1))$$

$$\stackrel{(d)}{=} \int_{v=1}^{v=0} \left(-\frac{d\omega(v)}{\omega(v)} + \frac{\omega(v)}{v^2} dv\right) - \log \omega(1)$$

$$= \left(-\log \omega(v) - \frac{1}{N} \sum_{n=1}^N \log\left(\frac{1}{v} + \frac{d_n^2}{\sigma^2}\right)\right)\bigg|_{v=1}^{v=0} - \log \omega(1)$$

$$\stackrel{(e)}{=} \frac{1}{N} \sum_{n=1}^N \log\left(1 + \frac{d_n^2}{\sigma^2}\right) \quad (A11)$$

where step (*a*) follows from Lemma 2, (*b*) from the matching condition (24), (*c*) by substituting $\rho = \phi(v)$, (*d*) from the fact that

$$d\phi(v) = -\frac{d\omega(v)}{\omega(v)^2} + \frac{dv}{v^2}$$

with 
$$\omega(v) = \frac{1}{v^{-1} + \phi(v)} = \frac{1}{N}\sum_{n=1}^{N}\left(\frac{1}{v} + \frac{d_n^2}{\sigma^2}\right)^{-1},$$

and (*e*) from the fact that

$$\lim_{v\to 0}\omega(v)\left(\prod_{n=1}^{N}\left(\frac{1}{v} + \frac{d_n^2}{\sigma^2}\right)\right)^{\frac{1}{N}} = 1.\qquad\blacksquare$$

## APPENDIX D
### PROOF OF THEOREM 2

Similarly to (20), $mmse(\rho)$ and $\psi(\rho)$ are related as
$$mmse(\rho) = \text{MMSE}(x_i' | \mathbf{b}') = \text{MMSE}(x_i' | \mathbf{b}_i', x_i' \sim p(x_i' | \mathbf{b}_{-i}'))$$
$$= \text{MMSE}(x_i' | \mathbf{b}_i', x_i' \sim \mathcal{CN}(\text{E}[x_i' | \mathbf{b}_{-i}'], \psi(\rho)))$$
$$= \gamma(\gamma^{-1}(\psi(\rho)) + \rho). \qquad (A12)$$

The above derivation mostly follows (20), except that, in the last step, we take into account the fact that $x_i'$ is constrained on a finite discrete constellation. Thus,

$$R \stackrel{(a)}{=} \int_0^\infty mmse(\rho)d\rho \stackrel{(b)}{=} \int_0^\infty \gamma(\gamma^{-1}(\psi(\rho)) + \rho)d\rho$$
$$\stackrel{(c)}{=} \int_0^{\phi(1)}\gamma(\rho)d\rho + \int_{\phi(1)}^{\phi(0)}\gamma(\gamma^{-1}(\phi^{-1}(\rho)) + \rho)d\rho$$
$$\stackrel{(d)}{=} \int_0^{\phi(1)}\gamma(\rho)d\rho + \int_{\phi(1)}^{\infty}\gamma(\rho + \rho')d(\rho + \rho') - \int_0^{\infty}\gamma(\rho' + \phi(\gamma(\rho')))d\rho'$$
$$\stackrel{(e)}{=} \log|\mathcal{S}| - \int_0^{\infty}\gamma(\rho + \phi(\gamma(\rho)))d\rho$$

where step (*a*) follows from Lemma 1 in [14], step (*b*) follows from (A12), step (*c*) follows from the curve-matching property in (24), step (*d*) follows by letting $\rho' = \gamma^{-1}(\phi^{-1}(\rho))$ and noting that $\rho' \in [0, \infty)$ as $\rho$ varies from $\phi(1)$ to $\phi(0)$, and step (*e*) utilizes the equality $\log|\mathcal{S}| = \int_0^\infty \gamma(\rho')d\rho'$ [14]. $\blacksquare$

## APPENDIX E
### PROOF OF LEMMA 3

It can be shown that the harmonic mean
$$f(g_1,...,g_N) = \left(N^{-1}\sum_{n=1}^{N}1/g_n\right)^{-1}$$
is concave in $\{g_n\}_{n=1}^{n=N}$ for $g_n > 0$, $1 \le n \le N$. Let
$$g_n = 1/\gamma(\rho) + \lambda_n^2 w_n / \sigma^2.$$
Thus, $\phi$ in (15) is concave in $\{w_n\}$. Together with the fact that $\gamma$ is convex and non-increasing (cf., [43] for the monotony of $\gamma$ in $\rho$), $\gamma(\rho + \phi(\gamma(\rho)))$ is convex in $\{w_n\}$ according to the composition rule of convex functions (cf., Chapter 3.2.4, [49]). Nonnegative weighted sums preserve convexity. Thus

$$\int_0^{+\infty}\gamma(\phi(\gamma(\rho)) + \rho)d\rho$$

is convex in $\{w_n\}$, which concludes the proof. $\blacksquare$

## APPENDIX F
### PROOF OF THEOREM 3

For a pair of matched detector and decoder, the maximum code rate is given by

$$R \stackrel{(a)}{=} \int_{v=1}^{v=0}\left(-\frac{d\omega(v)}{\omega(v)} + \frac{\omega(v)}{v^2}dv\right) - \log\omega(1)$$
$$= \left(-\log\omega(v) - \frac{1}{N}\text{E}\left[\log\det\left(\frac{1}{v}\mathbf{I} + \frac{1}{\sigma^2}\mathbf{H}\mathbf{Q}\mathbf{H}^{\text{H}}\right)\right]\right)\bigg|_{v=1}^{v=0} - \log\omega(1)$$
$$\stackrel{(b)}{=} \frac{1}{N}\text{E}\left[\log\det\left(\mathbf{I} + \frac{1}{\sigma^2}\mathbf{H}\mathbf{Q}\mathbf{H}^{\text{H}}\right)\right]$$

where step (*a*) follows the derivation in Appendix C (cf., step (d) of (A11)) except that $\omega(v)$ is now defined as

$$\omega(v) = \frac{1}{v^{-1} + \phi(v)} = \frac{1}{N}\text{tr}\left\{\text{E}\left[\left(\frac{1}{v}\mathbf{I} + \frac{1}{\sigma^2}\mathbf{H}\mathbf{Q}\mathbf{H}^{\text{H}}\right)^{-1}\right]\right\},$$

and step (*b*) is from the fact that

$$\lim_{v\to 0}\omega(v)\left(\text{E}\left[\det\left(\frac{1}{v}\mathbf{I} + \frac{1}{\sigma^2}\mathbf{H}\mathbf{Q}\mathbf{H}^{\text{H}}\right)\right]\right)^{\frac{1}{N}} = 1.\qquad\blacksquare$$